\documentclass[aps,prc,preprint,showpacs,preprintnumbers,amsmath,amssymb] {revtex4-1}
\usepackage{amsmath}
\usepackage{amssymb}
\usepackage{graphicx}
\usepackage{subfigure}
\usepackage{rotating}
\usepackage{natbib}
\usepackage{txfonts}
\usepackage{color}
\def\beq{\begin{equation}}
\def\eeq{\end{equation}}
\def\bea{\begin{eqnarray}}
\def\eea{\end{eqnarray}}

\def\eqref#1{Eq.~(\ref{eq:#1})}

\def\be{\begin{equation}}
\def\ee{\end{equation}}
\def\bg{\begin{eqnarray}}
\def\en{\end{eqnarray}}

\long\def\Omit#1{}

\DeclareGraphicsExtensions{ .pdf, .jpg, .eps}
\begin{document}
\title{${\bar p}p$ annihilation into ${\bar D} D$ meson pair within an effective 
Lagrangian model}
\author{R. Shyam$^{1,2}$}
\author{H. Lenske$^2$}
\affiliation{$^1$Saha Institute of Nuclear Physics, 1/AF Bidhan Nagar, Kolkata 700064, 
India}
\affiliation{$^2$Institut f\"ur Theoretische Physik, Universit\"at Gieseen, 
Heinrich-Buff-Ring 16, D-35392 Giessen, Germany} 

\date{\today}
\begin{abstract}
We study the charmed meson pair (${\bar D}^0 D^0$ and $D^+D^-$) production in 
${\bar p} p$ annihilation within an effective Lagrangian model that has only 
the baryon-meson degrees of freedom and involves the physical hadron masses. 
The reaction amplitudes include terms corresponding to the $t$-channel 
$\Lambda_c^+$, $\Sigma_c^+$, and $\Sigma_c^{++}$ baryon exchanges and the 
$s$-channel excitation, propagation and decay of the $\Psi(3770)$ resonance 
into the charmed mesons. The initial- and final-state distortion effects have 
been accounted for by using a simple eikonal approximation-based procedure in 
the same way as was done in our previous study of the ${\bar p} p \to 
{\bar \Lambda}_c^- \Lambda_c^+$ reaction within a similar model. The 
${\bar D}^0 D^0$ production reaction is dominated by the $\Lambda_c^+$ 
baryon exchange process and the corresponding total cross sections are 
predicted to be in the range of 0.18-0.7$\,\, \mu b$ for antiproton beam 
momenta varying between threshold and 20 GeV/c. The $\Psi(3770)$ resonance 
contributions have a large influence on the differential cross sections of 
the $D^- D^+$ production reaction.  
\end{abstract}
\pacs{13.60.Le, 14.40.Lb, 11.10.Ef,13.85.Fb}
\maketitle

\section{Introduction}

The first discovery of a charm-anticharm ($c{\bar c}$) bound state ($J/\psi$)
\cite{aub74,aug74} was made more than 30 years ago. Yet a substantial part 
of the charmonium spectrum is still to be precisely measured. Due to several 
reasons, charmonium states (and other heavy quarkonium states) have played an 
important role in our understanding of quantum chromodynamics (QCD), the 
fundamental theory of the strong interaction. Within the range of momentum 
exchange in bound $c{\bar c}$ systems, the value of the strong coupling constant 
$\alpha_s$ is not so large to invalidate the application of the perturbative 
methods. Thus these states provide a unique laboratory to explore the interplay 
between perturbative and nonperturbative effects in QCD. The relatively small 
binding energy of the charmonium as compared to the rest mass of its constituents 
allows its description by the nonrelativistic approaches that simplify and 
constrain the analysis of the nonperturbative effects (see, e.g. Ref.~\cite{bod12}, 
for a recent review). These mostly analytical methods are of considerable help in 
making significant progress in lattice QCD calculations, which have become 
increasingly more capable of dealing quantitatively with the nonperturbative 
dynamics in all its aspects (see, e.g., Refs.~\cite{kaw15,pre14}). 

Therefore, there is considerable interest in investigations of the production 
of charmonium states. Experimentally, they have been studied mainly in 
electron-positron and proton-antiproton annihilation processes. However, there 
are distinct advantages in producing $c{\bar c}$ states in the latter method 
where all the three valence quarks in a proton annihilate with their corresponding 
antiquark partners in an antiproton. This does not set any constraint on the quantum 
numbers of the final states enabling one to reach all the charmonium states by the 
direct formation. On the other hand, in electron-positron annihilation, the 
direct creation of final charmonium states is constrained to the quantum numbers 
of the photon ($J^{PC} =1^{--}$). Other states can be reached only indirectly by 
other mechanisms.  

The ${\bar P}ANDA$ ("antiproton annihilation at Darmstadt") experiment will use
the antiproton beam from the Facility for Antiproton and Ion Research (FAIR)
colliding with an internal proton target and a general purpose spectrometer to 
carry out a rich program on the charmonium production in proton-antiproton
annihilation. The entire energy region below and above the open charm threshold will
be explored in these studies. Charmonium states above the open charm threshold will
generally be identified by means of their decays to ${\bar D} D$
\cite{pan09,wie11,pre14a}, unless this is forbidden by some conservation rule.

The reliable estimation of the rates of ${\bar p} p \to {\bar D}^0 D^0$ and 
${\bar p} p \to D^- D^+$ reactions (to be together referred to as the ${\bar p} p 
\to {\bar D} D$ reaction) at the ${\bar P}ANDA$ energies is required for the
accurate detection of the charmonium states above the ${\bar D} D$ threshold. 
In addition, it is also important for other studies such as open charm 
spectroscopy, the search for charmed hybrids decaying to ${\bar D} D$, the 
investigation of the rare decays and of the charge-conjugation-parity  
violation in the $D$-meson sector. All these topics are the major components of 
the ${\bar P}ANDA$ physics program~\cite{pan09}. The accurate knowledge of these 
reactions is also the primary requirement for investigating the creation of the 
exotic flavored nuclear systems like charmed hypernuclei~\cite{dov77,tsu03,tsu04} 
and charmed $D$-mesic nuclei~\cite{tsu99,gar10,gar12}.  

The cross sections of the ${\bar p} p \to {\bar D} D$ reaction have been calculated 
by several authors employing a variety of models. In Ref.~\cite{kai94}, a 
nonperturbative approach has been used, which is based on the $1/N$ expansion in 
QCD, Regge asymptotics for hadron amplitudes, and a string model. Similar types 
of models were used in the calculations reported in Refs.~\cite{tit08,kho12}. In 
Ref.~\cite{gor13}, the ${\bar p} p \to {\bar D}^0 D^0$ reaction has been described 
within a double handbag model where the amplitude is calculated by convolutions of 
hard subprocess kernels (representing the transition $u{\bar u} \to c{\bar c}$ ) 
and the generalized parton distributions, which represent the soft nonperturbative 
physics. This approach was earlier used in Ref.~\cite{gor09} to describe the 
production of ${\bar \Lambda}_c^- \Lambda_c^+$ in ${\bar p} p$ annihilation and 
it resembles the quark-diquark picture that was employed by this group to make 
predictions for the cross sections of the $D^-D^+$ reaction in Ref.~\cite{kro89}. 

Recently the production of ${\bar D}D$ in antiproton-proton annihilation has been 
studied within the J\"ulich meson-exchange model in Ref.~\cite{hai14}. This 
approach was employed earlier to investigate the ${\bar p} p \to {\bar \Lambda} 
\Lambda$~\cite{hai92a,hai92b} and ${\bar p} p \to {\bar \Lambda}_c^- \Lambda_c^+$
\cite{hai10,hai11} reactions. In this model, these processes are considered within 
a coupled-channels framework, where the initial- and final-state interactions are 
taken into account in a rigorous way. The reactions proceed via the exchange of 
appropriate mesons between ${\bar p}$ and $p$ leading to the final baryon-antibaryon 
states.
  
Apart from Ref.~\cite{gor13}, where the calculated total cross section 
($\sigma_{tot}$) for the ${\bar p} p \to {\bar D}^0 D^0$ reaction was reported 
to be below 10 nb, in the majority of the calculations, the magnitudes of the 
$\sigma_{tot}$ for this reaction lie in the range of $10-100$ nb.  However, the 
predictions of various models differ significantly for the cross section of 
the ${\bar p} p \to  D^- D^+$ reaction. 
 
In this paper, we present the results of our investigations for the  cross 
sections of ${\bar p} p \to {\bar D}^0 D^0$, and $ {\bar p} p \to D^- D^+$ 
reactions within a single-channel effective Lagrangian model (see, e.g., Refs.
\cite{shy99,shy02,shy11}), where these reactions are described as a sum of the 
$t$-channel $\Lambda_c^+$, $\Sigma_c^+$, $\Sigma_c^{++}$ baryon exchange 
diagrams [see, Figs.~1(a) and 1(b)] and the $s$-channel excitation, propagation and 
decay into the $D{\bar D}$ channel of the $\Psi(3770)$ resonance (presented 
diagrammatically in Fig.~2). The $t$-channel part of the model is similar to that 
used in our previous calculation~\cite{shy14} of the ${\bar p} p \to {\bar 
\Lambda}_c^- \Lambda_c^+$  reaction that proceeds via the $t$-channel $D^0$ and 
$D^{*0}$ meson-exchange processes.   
\begin{figure}[t]
\centering
\includegraphics[width=.50\textwidth]{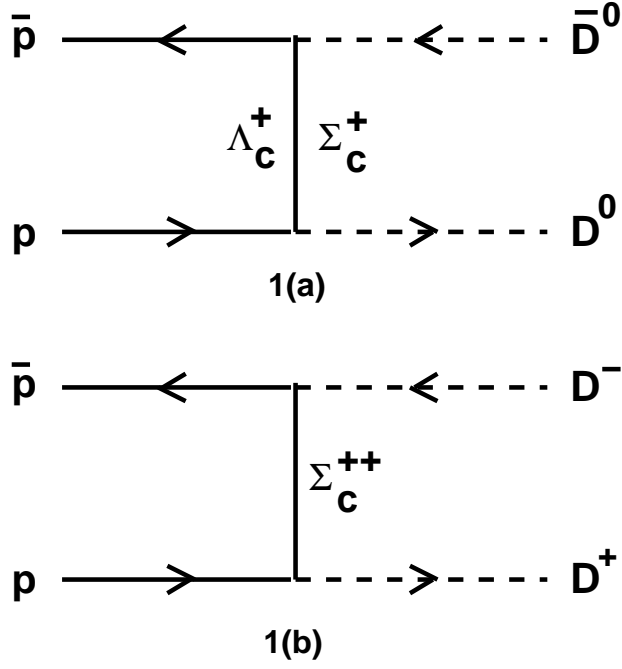}
\caption{
Graphical representation of the model used to describe the ${\bar p} + p \to
{\bar D}^0 + D^0$ (a) and ${\bar p} + p \to D^- + D^+$ (b) reactions via 
$t$-channel exchange of charmed baryons. In (a) $\Lambda_c^+$ and $\Sigma_c^+$ 
in the intermediate line represent the exchanges of $\Lambda_c^+$ and 
$\Sigma_c^+$ baryons, respectively while in (b) $\Sigma_c^{++}$ represents the 
exchange of $\Sigma_c^{++}$ baryon. 
}
\label{fig:Fig1}
\end{figure}

In the next section, we present our formalism. The results and discussions of our
work are given in Sec. III. Finally, the summary and the conclusions of this study
are presented in Sec. IV.  
 
\section{Formalism}

\begin{figure}[t]
\centering
\includegraphics[width=.50\textwidth]{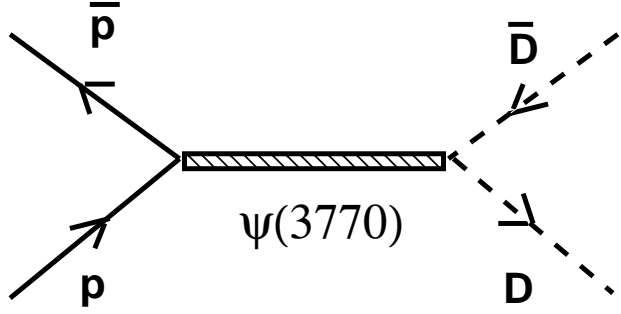}
\caption{
The Feynman diagram to describe the ${\bar p} + p \to {\bar D} + D$ reaction
via $s$-channel excitation, propagation and decay of the $\Psi(3770)$ resonance.  
}
\label{fig:Fig2}
\end{figure}

The exchanges of both $\Lambda_c^+$ and $\Sigma_c^+$ baryons contribute to the 
$t$-channel amplitude of the ${\bar p} + p \to {\bar D}^0 + D^0$ reaction. However, 
the process ${\bar p} + p \to D^- + D^+$ is mediated only by the exchange of the
$\Sigma_c^{++}$ baryon in the $t$-channel. On the other hand, the $s$-channel 
excitation, propagation and subsequent decay of the intermediate resonance state 
$\psi(3770)$ contribute to both these reactions. 

To evaluate amplitudes for the processes shown in Figs.~1 and 2, we have used the
effective Lagrangians at the charm baryon-meson-nucleon vertices, which are taken 
from Refs.~\cite{shy14,hob14,hai11a,gro90,wie10}. For the vertices involved in the 
$t$-channel diagrams we have 

\begin{eqnarray}
{\cal L}_{NBD} & = & ig_{NBD}{\bar \psi}_N i\gamma^5 \phi_{D}\psi_B + H.c.,
\end{eqnarray}
\noindent
where ${\psi}_N$ and $\psi_B$ are the nucleon (antinucleon) and charmed baryon
fields, respectively, and $\phi_{D}$ is the $D$-meson field. $g_{NBD}$
in Eq.~(1) represents the vertex coupling constant.

The amplitude of the diagrams given in Figs. 1(a) and 1(b) is given by

\begin{eqnarray}
A(B) & = & i \frac{g_{NBD}^2}{q^2-(m_B-i\Gamma_B/2)^2}\,\,{\bar \psi}_{\bar p}
(k_{\bar p}) \,\gamma^5 \, (\gamma_\mu q^\mu + m_B)\, \gamma^5\, \psi_p(k_p),
\end{eqnarray}
\noindent
where $B$ represents the exchanged charmed baryon. $q$, $m_B$ and $\Gamma_B$ are 
the momentum, mass and the width of the exchanged charmed baryon, respectively. 
The term that contains these quantities comes from the propagator of these baryons. 
The widths of the charmed baryons are taken from the latest Particle Data Group 
estimates~\cite{oli14}. The coupling constants $g_{NBD}$ are adopted from Refs.
~\cite{hai11a,hob14}, as $g_{N \Lambda_c^+ D}$ = 13.50, $g_{N \Sigma_c^+ D}$ = 
2.69 and $g_{N \Sigma_c^{++} D}$ = 2.69.  From these values it is expected that 
$\Lambda_c^+$ will dominate the $t$-channel production amplitudes.
 
The off-shell behavior of the vertices is regulated by a monopole form factor
(see, e.g., Refs.~\cite{shy99,shy02})
\begin{eqnarray}
F_i(q_{B_i}) & = & \frac{\lambda_i^2-m_{B_i}^2}{\lambda_i^2-q_{B_i}^2},
\end{eqnarray}
where $q_{B_i}$ is the momentum of the {\it i}th exchanged baryon with mass 
$m_{B_i}$. $\lambda_i$ is the corresponding cutoff parameter, which governs 
the range of suppression of the contributions of high momenta carried out via 
the form factor. We chose a value of 3.0 GeV for $\lambda_i$ at all the vertices. 
The same $\lambda_i$ was also used in the monopole form factor employed in the 
study of the reaction ${\bar p} p \to {\bar \Lambda}_c^- \Lambda_c^+$ in Ref.
\cite{shy14} within a similar model. It may be mentioned here that in the 
J\"ulich meson-exchange model calculations of the reaction ${\bar p} + p \to 
{\bar D} + D$ presented in Ref.~\cite{hai14}, a form factor of the following type
has been used:
\begin{eqnarray}
F_{i}(q_{B_i}) & = & \left [\frac{{(\lambda_i)^4}}
{{(\lambda_i)^4} + (q_i^2-m_{B_i}^2)^2} \right ]
\end{eqnarray}
with a $\lambda_i$ of 3.5 GeV. This form factor gives more weight to the lower 
momentum transfers. As discussed below we use this type of form factor at 
the resonance production and decay vertices.  

In order to evaluate the diagram of Fig.~2, the effective Lagrangians are required 
at $\Psi {\bar p} p$ and $\Psi {\bar D} D$ vertices, which are written as
\begin{eqnarray}
{\cal L}_\mu^{\Psi {\bar p} p} & = & g_{\Psi {\bar p} p}\, \Big[{\bar \psi}_{\bar p}
\,\big(\gamma_\mu + \frac{\kappa_\Psi}{2M}\sigma_{\mu \nu}\,\partial^\nu \,
\theta_\Psi^\mu \big) \psi_p \Big],
\end{eqnarray}
and
\begin{eqnarray}
{\cal L}_{\Psi {\bar D} D} & = & g_{\Psi {\bar D} D} \,\big(\Phi_{\bar D}\, 
\partial_\mu \,\Phi_D \big) \, \theta_\Psi^\mu.
\end{eqnarray}
\noindent
In Eq.~(5) $M$ represents the nucleon mass and $\theta_\Psi^\mu$ is the 
$\Psi$ resonance field. $g_{\Psi {\bar p} p}$ and $\kappa_\Psi$ are the coupling 
constants at the $\Psi {\bar p} p$ vertex. Similarly, in Eq.~(6) 
$g_{\Psi {\bar D} D}$ is the coupling constant at the $\Psi {\bar D} D$ vertex and
$\Phi_{\bar D}$ and $\Phi_D$ represent the ${\bar D}$ and $D$ charmed meson fields.
The values of the coupling constants $g_{\Psi {\bar D}^0 D^0}$, $g_{\Psi D^- D^+}$ 
and $g_{\Psi {\bar p} p}$ have been determined from the branching ratios for the 
decay of $\Psi(3770)$ resonance to the relevant channels as given in Refs.
\cite{abl06} and~\cite{abl14}. We take $g_{\Psi {\bar D}^0 D^0}$ = 17.90 (see also
Ref.~\cite{che13}), $g_{\Psi D^- D^+}$ = 14.10 and $g_{\Psi {\bar p} p}$ = 
5.12$\times$10$^{-3}$. The value of $\kappa_\Psi$ is fixed by fitting the 
cross sections of the ${\bar p} + p \to \Psi(3770) \to {\bar D}^0 D^0$ calculated 
within the effective Lagrangian model to that obtained within a semiclassical
resonance production and decay model where experimental widths for the decay
processes $\Psi \to {\bar p} p$ and $\Psi \to {\bar D} D$ are used. This is 
discussed in the next section.

The amplitude of the process ${\bar p} + p \to \Psi(3770) \to {\bar D} D$
(Fig.~2) is written as
 
\begin{eqnarray}
A(\Psi) & = & -g_{\Psi {\bar p} p}g_{\Psi {\bar D} D}\, \frac{1}
{s_{inv}-(m_\Psi-i\Gamma_\Psi/2)^2} \,\Big[\psi_{{\bar p}}\big(\gamma_\mu +
\frac{i\kappa_\Psi} {2M}\,\sigma_{\mu \nu}\,q^\nu \big)\psi_p\Big]
\,(k_{\bar D}-k_{D})^\mu,
\end{eqnarray}
\noindent
where $k_{\bar D}$ and $k_{D}$ are the momenta associated with the final-state 
${\bar D}$ and $D$ mesons, respectively, and $s_{inv}$ is the square of the 
invariant mass associated with the $\Psi$ resonance. It may be mentioned that 
the denominator of the $\Psi$ propagator leads to a cross section that has a 
pole in the vicinity of the physical mass value ($m_\Psi$) of the $\Psi$ 
resonance, which is taken to be 3773.15 MeV. The total width ($\Gamma_\Psi$) of 
this resonance is 27.2 MeV~\cite{oli14}. A similar approach for the denominator 
of the $\Psi$ propagator has also been adopted in Refs.~\cite{zha10,lim14,hai15}. 
This procedure is inspired by the extreme vector-meson dominance hypothesis 
where contributions of the vector-meson resonance are included in the vicinity 
of the relevant kinematical regime and in the far off-shell vector-meson 
kinematical regions it is considered  as background that is generally quite
weak. 

In order to account for the off-shell effects due to the internal structure of 
the intermediate charmonium states, we introduce vertex form factors for the 
$\Psi$ resonance. In our calculations the shapes of these form factors are 
given by Eq.~(4) with cutoff parameters  
\begin{eqnarray}
\lambda_\Psi = m_\Psi + \alpha \lambda_{QCD},
\end{eqnarray}
where $\lambda_{QCD}$ = 240 MeV, and $\alpha$ is an adjustable parameter. We 
have used $\alpha$ = 7.5 for amplitudes involving the $\Psi(3770)$ resonance. 

A widely used approach is to parametrize the  total cross section for the 
process ${\bar p} p \to \psi(3770) \to {\bar D} D$  in a Breit-Wigner form 
(see, e.g. Refs.~\cite{abl06,tei97,ach12})

\begin{eqnarray}
\sigma_{{\bar p}p \to \psi(3770) \to {\bar D} D} & = & 
\frac{2J_\Psi + 1}{(2j_{\bar p}+1)(2j_p+1)}\, \frac {4\pi}{q_{{\bar p}p}^2}\,\, 
\frac{s_{inv}\Gamma_{\Psi(3770) \to {\bar p}p}\,\,\Gamma_{\Psi(3770) \to {\bar D} D}}
{(s_{inv}-M_\psi^2)^2 + s_{inv}\Gamma_{tot}^2},
\end{eqnarray} 
\noindent
where $J_\Psi$ is the spin of the resonance and $q_{{\bar p}p}$ is the momentum in 
${\bar p}p$ channel. $\Gamma_{\Psi(3770) \to {\bar p} p}$ is the partial 
width for the production process ${\bar p} p \to \Psi$ and $\Gamma_{\Psi(3770) 
\to {\bar D} D}$ is the partial width for the $\Psi \to {\bar D} D$ decay.
$\Gamma_{tot}$ is the total width of the $\Psi$ resonance. Attempting to account 
for the possible self-energy contributions to the formation and decay of the 
resonance, some authors (see, e.g., Ref.~\cite{ach12}) have introduced an energy 
dependence to the width $\Gamma_{\Psi(3770) \to {\bar D} D}$ adopted from Ref.
\cite{bla52}. The ansatz for this energy dependence involves the range of ${\bar D}
D$ interactions, which is treated as a free fitting parameter. For the sake of 
simplicity and in order to keep the number of adjustable parameters as small as 
possible, here we refrain from such a  more elaborate approach. We use a constant 
width, which is a good approximation in view of the very narrow line width of the 
$\Psi(3770)$ resonance. In fact, close to the resonance pole, the expression given 
by Eq.~(9) is indeed a very good approximation of the exact result~\cite{hai15}. 

In our numerical calculations, we have used for $\Gamma_{\Psi(3770) \to {\bar D} D}
$ the values extracted from the  branching ratios of this decay as given in the 
latest compilation of the Particle Data Group (PDG) group~\cite{oli14}. The width 
$\Gamma_{\Psi(3770)} \to {\bar p}p$ is obtained from the branching fraction 
$B_{\Psi \to {\bar p} p }$ = $7.1_{-2.9}^{+8.6} \times 10^{-6}$ as reported in 
Ref.~\cite{abl14}. We have taken the width corresponding to the upper limit of 
$B_{\Psi \to {\bar p} p }$, which was also employed in the determination of the 
coupling constant $g_{\Psi {\bar p} p}$ used in Eq.~(7).  

From the studies of the ${\bar \Lambda}_c^- \Lambda_c^+$ production
\cite{hai10,shy14}, it is well known that the magnitudes of the cross sections 
depend very sensitively on the initial-state distortion effects. In fact, the 
${\bar p} p$ annihilation channel is almost as strong as the elastic scattering
channel. This large depletion of the flux can be accounted for by introducing 
absorptive potentials that are used in optical models or in coupled-channels 
approaches~\cite{koh86,hai92a,hai92b,alb93,hai10}. In this work, instead of 
employing such a detailed treatment, we use a procedure that was originated by 
Sopkovich~\cite{sop62} and was employed in Ref.~\cite{shy14} for describing the  
${\bar p} p \to {\bar \Lambda}_c^- \Lambda_c^+$ reaction. In this method, the 
transition amplitude for the reaction ${\bar p} p \to {\bar D} D $ with distortion 
effects is written as

\begin{eqnarray}
T^{{\bar p}p \to {\bar D} D} & = &
\sqrt{\Omega^{{\bar p}p}}T^{{\bar p}p \to {\bar D} D}_{Born}
\sqrt{\Omega^{{\bar D} D}} 
\end{eqnarray}
\noindent
where $T^{{\bar p}p \to {\bar D} D}_{Born}$ is the transition matrix calculated 
within the plane-wave approximation and $\Omega^{{\bar p}p}$ and the 
$\Omega^{{\bar D} D}$ are the operators describing the initial- and final-state 
elastic interactions, respectively.  

For the present purpose, we neglect the real part of the baryon-antibaryon 
interaction. Considering the ${\bar p}p$ initial-state interaction (ISI), we 
describe the strong absorption by an imaginary potential of Gaussian shape with 
range parameter $\mu$ and strength $V_0$. By using the eikonal approximation, the 
corresponding attenuation integral can be evaluated in a closed form. Similar to 
Refs.~\cite{sop62,rob91}, we obtain for $\Omega^{{\bar p}p}$ 
 
\begin{eqnarray}
\Omega^{{\bar p}p} & = & exp\big[\frac{-\sqrt{\pi}EV_0}{\mu k} exp(-\mu^2 b^2)\big],
\end{eqnarray} 
\noindent
where $b$ is the impact parameter of the ${\bar p}p$ collision. $E$ and $k$ are the 
center-of-mass energy and the momentum of the particular channel, respectively.
In our numerical calculations, we have used the same values for the parameters 
$V_0$, $\mu$ and $b$ as in Ref.~\cite{shy14}. It may be noted that with these 
parameters we were able to get cross sections for the ${\bar p}p \to {\bar \Lambda}
\Lambda$ strangeness production reaction in close agreement with the corresponding 
experimental data.  Furthermore, our cross sections for the  ${\bar p} p \to {\bar 
\Lambda}_c^- \Lambda_c^+$ reaction were similar in magnitude to those reported in a 
coupled-channels meson-exchange model calculation~\cite{hai10}.

In the case of the ${\bar p} p \to {\bar \Lambda}_c^- \Lambda_c^+$ reaction, it 
has been shown in Ref.~\cite{hai10} that, because of the strong absorption in the 
initial channel, the production cross sections were rather insensitive to the 
final-state interaction (FSI) between ${\bar \Lambda}_c^-$ and $\Lambda_c^+$. In 
Ref.~\cite{hai14}, the effects of ${\bar D}D$ FSI were investigated by approximately 
extending in the charmed meson sector the $\pi \pi \to {\bar K}K$ model of the 
J\"ulich group~\cite{loh90}. However, these calculations have sizable uncertainties 
and even then the effect of FSI is not big. In our procedure, unlike the ${\bar p}p$ 
ISI, it is not possible to put any constraint, experimental or otherwise, on the 
choice of the ${\bar D}D$ FSI distortion parameters. Therefore, in order to keep 
the number of free parameters small, like our study on ${\bar \Lambda}_c^- 
\Lambda_c^+$ production, we concentrate only on the initial-state interaction in 
this study. It should be mentioned that also in calculations reported in Refs.
\cite{kho12,kai94,tit08} the meson-meson FSI effects were not considered.
\begin{figure}[t]
\centering
\includegraphics[width=.50\textwidth]{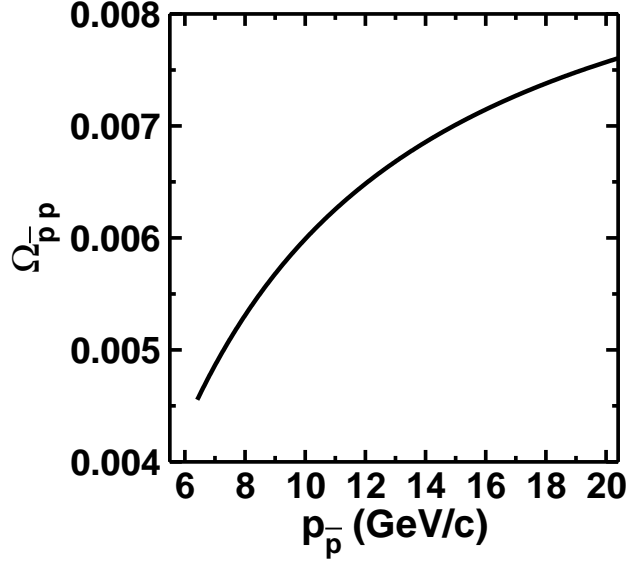}
\caption{
The distortion factor $\Omega_{{\bar p} p}$  [defined by Eq. (11)] as a function 
of the antiproton beam momentum. The values of parameters $\mu$, $V_0$ and $b$
were taken to be 0.3369 GeV, 0.8965 GeV and 0.3270 GeV$^{-1}$, respectively,
which are the same those used in Ref.~\cite{shy14}.  
}
\label{fig:Fig3}
\end{figure}

The distortion effects could lead to the reduction of the undistorted cross 
sections by several orders of magnitude depending upon the values of the parameters 
$V_0$, $\mu$ and $b$. In Fig.~3, we have shown $\Omega_{\bar p p}$ as function 
of the beam momentum $p_{\bar p}$ with the values of parameters $\mu$, $V_0$ and $b$ 
being 0.3369 GeV, 0.8965 GeV and 0.3270 GeV$^{-1}$, respectively. We see that 
$\Omega_{\bar p p}$ increases gradually as $p_{\bar p}$ goes beyond the threshold and 
becomes almost constant at higher values of $p_{\bar p}$ $-$ in this region the 
values of $E$ and $k$ are roughly  equal. With this $\Omega_{\bar p p}$, the 
undistorted cross sections would dampen by almost 2 orders of magnitude. As far as 
dependencies on the parameters are concerned, increasing $V_0$ obviously increases 
the damping, while decreasing $\mu$ has the same effect. 

\section{Results and Discussions}

One can estimate the total cross sections for the ${\bar p}p \to {\bar D}^0 D^0$ 
and ${\bar p}p \to D^- D^+$ reactions around the $\Psi(3770)$ peak with the help
of Eq.~(9). With the values of various widths as extracted from the experimental 
data as specified in the last section, the total cross sections calculated by 
Eq.~(9) can indeed be used to fix some of the parameters of the effective 
Lagrangian model (ELM). We deduce the parameter $\kappa_\Psi$ in Eq.~{5} by 
comparing the total cross section for e.g., the ${\bar p} p \to \psi(3770) \to 
{\bar D}^0 D^0$ reaction calculated within the ELM [by using the amplitude given 
by Eq.~(7)] with that obtained by Eq.~(9). In the ELM calculations the ${\bar p}p$ 
ISI has been included. In Fig.~3, we show this comparison where the value of 
$\kappa_\Psi$ is taken to be 6.0.
\begin{figure}[t]
\centering
\includegraphics[width=.50\textwidth]{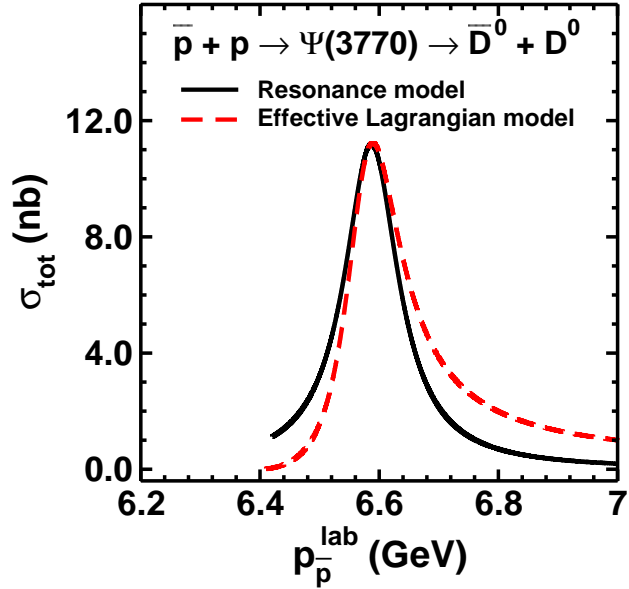}
\caption{
Total cross section for the ${\bar p} p \to \Psi(3770) \to {\bar D}^0 D^0$ 
reaction as a function of the antiproton beam momentum. The solid line represents 
the cross sections calculated within the resonance production and decay model 
[Eq.~(9)], while the dashed line shows the effective Lagrangian model cross 
sections [obtained with the amplitude given by Eq.~(7)]. 
}
\label{fig:Fig4}
\end{figure}

In Fig.~4, we see that the peak of the ELM cross sections coincides with that 
of the resonance model. Also, in the vicinity of the resonance peak the two models 
predict the same magnitudes of the cross sections and the same line shapes.  
The value of this cross section is about 12 nb at the peak position, which is close 
to the lower limit of $\sigma_{tot}$ predicted in Ref.~\cite{hai15}.

In Figs.~5(a) and 5(b) we present the results for the total cross sections of
${\bar p} p \to {\bar D}^0 D^0$ and ${\bar p} p \to D^- D^+$ reactions, 
respectively, for antiproton beam momenta ($p_{{\bar p}}^{lab}$) in the  region 
of $6.4 - 6.8$ GeV/c, where the $\Psi(3770)$ resonance is expected to show its 
impact. In these figures, solid lines represent the cross sections where the 
amplitudes of the $t$-channel baryon exchanges and the $s$-channel $\Psi(3770)$ 
production and decay are coherently added, while the dashed lines include only 
the $t$-channel baryon exchange contributions.
\begin{figure}[t]
\centering
\includegraphics[width=.50\textwidth]{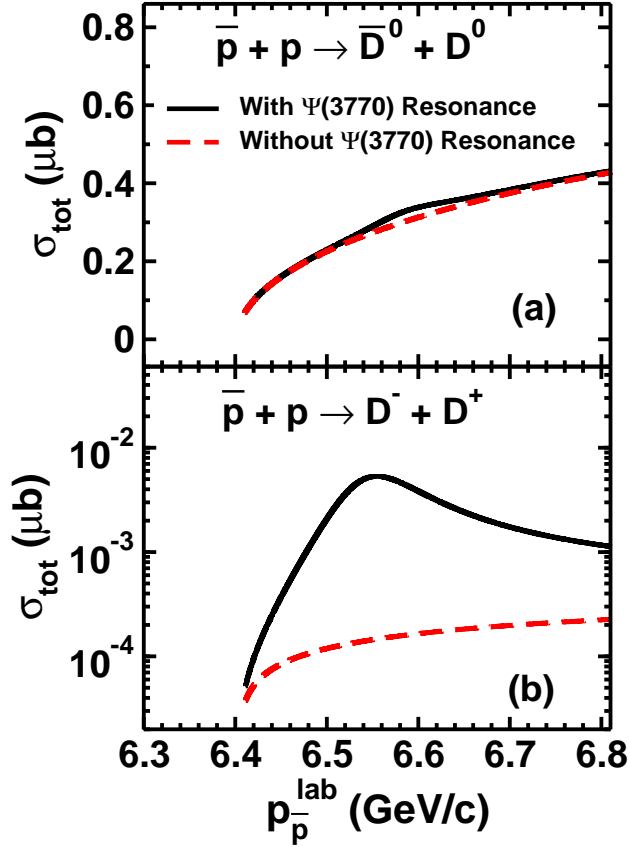}
\caption{
(a) Total cross section for the ${\bar p} p \to {\bar D}^0 D^0$ reaction as a 
function of the antiproton beam momentum. The solid and dashed lines represent 
the cross sections obtained by the coherent sum of the $t$-channel baryonic 
exchange and the $s$-channel $\Psi(3770)$ resonance excitation amplitudes, and 
by including the $t$-channel baryon exchange contributions only in the amplitude, 
respectively. (b) The same for the ${\bar p} p \to D^- D^+$ reaction as a function 
of the antiproton beam momentum.  The full and dashed lines have the same meaning 
as in (a).
 }
\label{fig:Fig5}
\end{figure}
We see in Fig.~5(a) that for the ${\bar p}p \to {\bar D}^0 D^0$ reaction the 
cross sections corresponding to the $t$-channel baryon exchange processes are 
fairly large ($\sim$ 300 nb) in the resonance peak (RP) region, so the inclusion 
of the $\Psi(3770)$ resonance contributions does not make any dramatic effect. 
It just produces a small kink in the cross section around the RP momentum. This 
is in contrast to the results of Ref.~\cite{hai15}, where the $\Psi(3770)$ 
resonance led to an enhancement of almost a factor of two in the cross sections 
of the ${\bar p}p \to {\bar D}^0 D^0$ reaction around the RP region. The reasons 
for this difference can be attributed to two facts. First, our $t$-channel 
baryon exchange cross sections are larger than those of Ref.~\cite{hai15} (about 
300 nb as compared to only about 40 nb) in the RP region, and second, our cross 
sections for the $\Psi(3770)$ resonance excitation are smaller than those of Ref.
\cite{hai15} in this region. 

On the other hand, in our study the effect of the $\Psi(3770)$ resonance is quite 
prominent for the  ${\bar p}p \to D^- D^+$ reaction as can be seen in Fig.~5(b). 
In our model, the $t$-channel baryon exchange contributions to the cross sections 
of this reaction are strongly suppressed. The reason for this is that only the 
$\Sigma^{++}$ exchange mediates the $t$-channel amplitude in this case, which 
becomes very small because of the much smaller coupling constant and somewhat larger 
mass of the exchanged baryon. The ratio of the absolute magnitudes of the ${\bar p}p 
\to {\bar D}^0 D^0$ and ${\bar p}p \to D^- D^+$ reactions is roughly proportional 
to $(g_{N\Lambda_c^+D}/g_{N\Sigma_c^{++}D})^4$, which leads to a reduction in the 
$D^-D^+$ production cross section over that of ${\bar D}^0 D^0$ by nearly a factor 
of 650. 

It should, however, be mentioned here that in the consideration of the ISI within 
the distorted-wave Born approximation (DWBA) approach [Ref.~\cite{hai14}], two-step 
transitions of the form ${\bar p}p \to {\bar n}n \to D^-D^+$, are generated. Because 
the $\Lambda_C^+$ exchange can contribute to the ${\bar n}n \to D^-D^+$ transition 
potential, this exchange is no longer absent. Therefore, these two-step mechanisms 
can enhance the $D^-D^+$ production cross sections. Indeed, in Refs.
\cite{hai14,hai15} the cross sections for $D^-D^+$ production are even larger than 
those of the ${\bar D}^0 D^0$ production. On the other hand, such two-step mechanisms 
are out of the scope of our as well as of Regge model~\cite{kai94,tit08,kho12} 
calculations.  Therefore, in studies within these models the cross sections of the 
${\bar p}p \to D^-D^+$ reaction are suppressed as compared to those of the 
${\bar p}p \to {\bar D}^0 D^0$ reaction.
\begin{figure}[t]
\centering
\includegraphics[width=.50\textwidth]{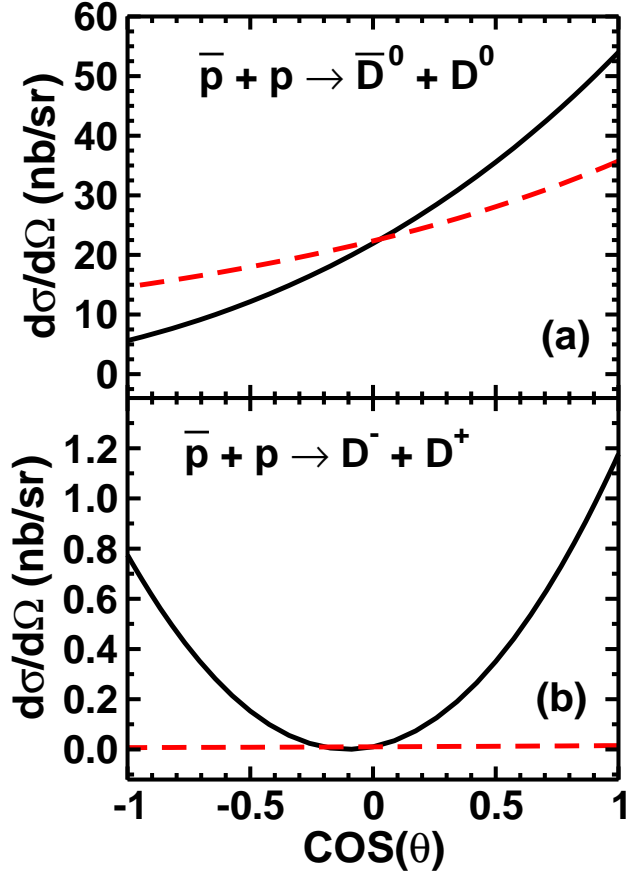}
\caption{
(a) Differential cross section for the ${\bar p} p \to {\bar D}^0 D^0$ reaction at 
the antiproton beam momentum of 6.57 GeV/c. The solid and dashed lines have the 
same meanings as those in Fig.~5. (b) The same as in (a) but for the ${\bar p} p 
\to  D^- D^+$ reaction.
}
\label{fig:Fig6}
\end{figure}

In Figs.~6(a) and 6(b) we show the differential cross sections (DCS) for ${\bar p}p 
\to {\bar D}^0 D^0$ and ${\bar p}p \to D^- D^+$ reactions, respectively, at the 
beam momentum of 6.57 GeV/c, which corresponds to the $\Psi$ resonance invariant 
mass ($\sqrt{s_{inv}}$) = 3770.24 MeV. Therefore, these cross sections represent 
the angular distributions of the produced charmed mesons at practically the 
resonance peak. In Fig.~6(a), we note that the inclusion of the $\Psi(3770)$ 
resonance alters significantly the cross section obtained with only the $t$-channel 
baryon exchange term (dashed line). Strong interference effects of various amplitudes 
are evident. At the backward angles, the $t$-channel baryon exchange and $s$-channel 
$\Psi(3770)$ amplitudes interfere destructively while at forward angles this 
interference is constructive leading to the strong forward peaking of the angular 
distribution. Because of this interference effect, the DCS do not exhibit the 
type of shape that is expected from a pure $p$-wave resonance dominated amplitude.

On the other hand, in Fig.~6(b), the differential cross section obtained by 
adding the $s$-channel $\Psi(3770)$ terms to the $t$-channel baryon exchange 
amplitudes shows a dominant $p$-wave type of angular distribution. This is due 
to the fact that contributions of the $\psi(3770)$ resonance term are significantly
stronger than those of the $t$-channel baryon exchange term in this case. This was
apparent already in Fig.~5(b). However, even though the $t$-channel baryon exchange 
amplitudes are relatively quite small, they do introduce some distortion to the 
angular distribution of the $D$ mesons arising from the decay of the $\Psi(3770)$ 
resonance through the interference terms. This is evident from the asymmetry of 
the cos($\theta$) distribution depicted by the solid curve.
\begin{figure}[t]
\centering
\includegraphics[width=.50\textwidth]{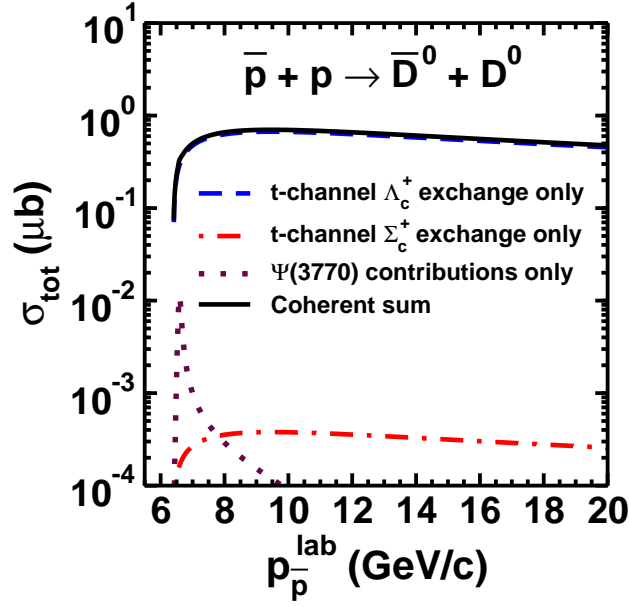}
\caption{
Total cross section for the ${\bar p} p \to {\bar D}^0 D^0$ reaction as a function 
of the antiproton beam momentum. The contributions of $t$-channel $\Lambda_c^+$ and
$\Sigma_c^+$ baryon exchange processes are shown by dashed and dashed-dotted lines,
respectively, while the dotted line represents the cross sections of the $s$-channel 
$\Psi(3770)$ resonance excitation. The cross sections corresponding to the coherent 
sum of these amplitudes are shown by the full curve.
}
\label{fig:Fig7}
\end{figure}

In general, the interference patterns in the differential cross sections depend 
quite sensitively on the relative magnitudes of the $t$-channel baryon exchange 
and $s$-channel $\Psi(3770)$ amplitudes. Thus, they provide a critical check on 
the coupling constants that enter into these amplitudes. Therefore, measurements 
of the differential cross sections of these reactions in future experiments would 
be useful in fixing these coupling constants. 

Next, we discuss the charm meson production in antiproton-proton annihilation at
higher beam energies. In Fig.~7, we show the total cross section of the ${\bar p} 
+ p \to {\bar D}^0 D^0$ reaction for antiproton beam momentum varying in the range 
of threshold to 20 GeV/c. In this figure, the roles of various $t$-channel baryon 
exchange and the $s$-channel $\Psi(3770)$ resonance excitation processes have been 
investigated. We note that $\sigma_{tot}$ for this case is almost solely governed by 
the $\Lambda_c$-exchange mechanism in the entire range of the antiproton beam momentum. 
The contributions of $\Sigma_c^+$-exchange terms are lower by about 3 orders of 
magnitudes. This can be understood from the approximate proportionality of the ratio 
of $\sigma_{tot}$ of the two exchange processes to the fourth power of the ratio of 
the coupling constants of respective vertices involved in the corresponding amplitudes 
as discussed above. The $s$-channel $\Psi(3770)$-exchange term contributes negligibly 
to the $\sigma_{tot}$ of this reaction at higher beam momenta.  

We further note in Fig.~7 that the $\sigma_{tot}$ peaks at $p_{{\bar p}}^{lab}$
of about 9 GeV/c. This is in agreement with the results of the Regge trajectory 
model calculations of Refs.~\cite{kho12} and \cite{kai94}. At the beam momentum of 
interest to the ${\bar P}ANDA$ experiment (15 GeV/c), total cross section for the 
${\bar D}^0 D^0$ production reaction predicted by our model is about 570 nb. This 
should be compared with the results reported by other authors for this beam momentum.  
In Refs.~\cite{kho12} and ~\cite{kai94} the corresponding cross sections are 
approximately 100 and 70 nb, respectively, while in Ref.~\cite{gor13} it is less 
than 10 nb. We recall that in Refs.~\cite{kho12} and \cite{kai94}, the ISI effects 
were included by following an eikonal-model-based procedure similar to that of our 
study.  Of course, uncertainties in this method can not be  ruled out. However, in our 
calculations, we have taken the same ISI parameters in Eq.~(11) as those used in our 
previous study~\cite{shy14} of the ${\bar p} p \to {\bar \Lambda}_c^- \Lambda_c^+$ 
reaction within a similar model. These parameters were checked  by reproducing the 
near-threshold cross section predicted within the J\"ulich meson-exchange model 
calculations of this reaction reported in Refs.~\cite{hai10,hai11} where ISI effects 
have been treated more rigorously within a coupled-channels method.
\begin{figure}[t]
\centering
\includegraphics[width=.50\textwidth]{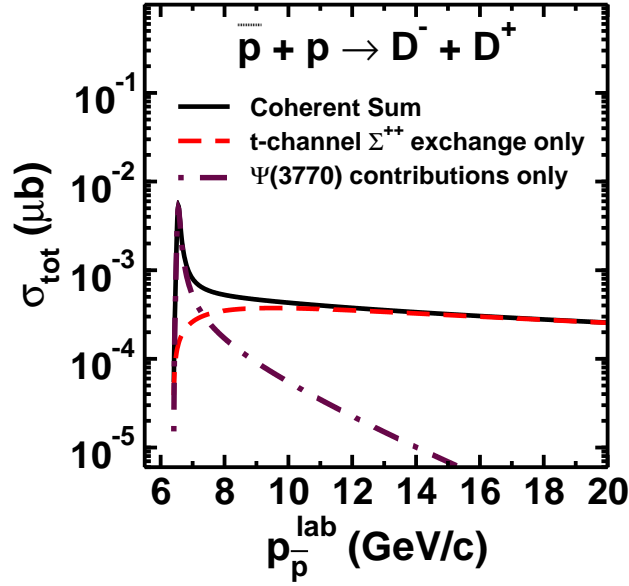}
\caption{
Total cross section for the ${\bar p} p \to D^- D^+$ reaction as a function
of the antiproton beam momentum. The contributions of $t$-channel $\Sigma_c^{++}$ 
baryon exchange and $s$-channel $\Psi(3770)$ resonance excitation are shown by 
dashed and dashed-dotted lines respectively. The cross sections corresponding to 
the coherent sum of these terms in the total amplitudes are shown by the full curve.
}
\label{fig:Fig8}
\end{figure}

In Fig.~8, we present the total cross sections for the ${\bar p} p \to D^- D^+$ 
reaction as a function of the antiproton beam momentum. We see that the 
$\sigma_{tot}$ of this reaction is strongly suppressed compared to that of Fig.~7. 
This was seen already in Fig.~5 at near-threshold beam momenta. A similar 
suppression of $D^-D^+$ cross sections relative to those of ${\bar D}^0 D^0$ has 
been noted in Refs.~\cite{kho12}, \cite{tit08}, and \cite{kai94}. This can be 
attributed to the much smaller $\Sigma_c^{++}$-exchange vertex coupling constant in 
comparison to that of the $\Lambda_c^+$-exchange vertex. However, in the 
coupled-channels meson-exchange model, the initial-state inelastic interactions 
could enhance the $D^-D^+$ cross sections significantly as was discussed earlier. 
It would be quite interesting to test the predictions of various models for 
the $D^-D^+$ cross sections in the ${\bar P}ANDA$ experiment at the upcoming FAIR 
facility. 
\begin{figure}[t]
\centering
\includegraphics[width=.50\textwidth]{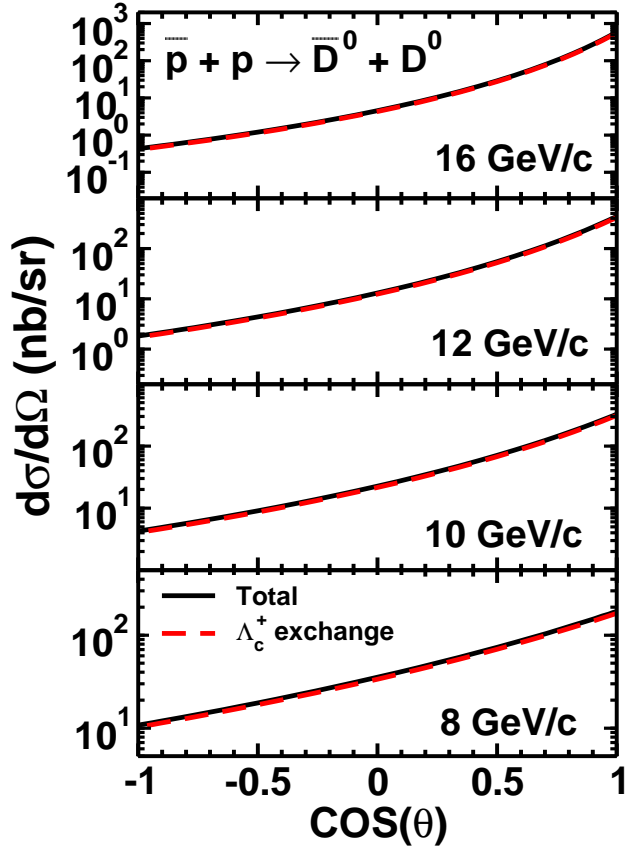}
\caption{
Differential cross section for the ${\bar p} p \to {\bar D}^0 D^0$ reaction at 
the antiproton beam momenta of 8, 10, 12 and 16 GeV/c as indicated in the figure. 
The solid lines show the cross sections where the amplitudes of the $t$-channel 
$\Lambda_c^+$, and $\Sigma_c^+$ baryon exchange and $s$-channel $\Psi(3770)$ 
resonance terms are coherently added. The dashed lines show the cross sections 
where the amplitudes include the contributions of the $\lambda_c^+$-exchange term 
only. 
}
\label{fig:Fig9}
\end{figure}

As was already noted in Figs. 6(a) and 6(b), the differential cross sections 
provide more explicit information about the reaction mechanism. These cross 
sections involve terms that weigh the interference terms of various components of 
the amplitude with the angles of the outgoing particles. Therefore, in general 
the contributions of different mechanisms are highlighted in different angular 
regions. In Fig.~9, we show the predictions of our model for the differential 
cross sections for the ${\bar p} p \to {\bar D}^0 D^0$ reaction at the beam momenta 
of 8, 10, 12 and 16 GeV/c. In this figure, the solid lines represent cross 
sections that include the coherent sum of the amplitudes corresponding to the 
$t$-channel $\Lambda_c^+$ and $\Sigma_c^+$ baryon exchanges and the $s$-channel 
$\Psi(3770)$ resonance terms, while the dashed lines show the cross sections where 
the amplitudes include contributions of the $\Lambda_c^+$-exchange term only. Since,
the $\Sigma_c^+$ baryon exchange and the $\Psi(3770)$ resonance contributions are 
quite small compared to those of the $\Lambda_c^+$ exchange, the interference 
effects of various terms in the amplitudes are not significant at higher beam 
momenta. We notice that with increasing beam momentum cross sections are more and 
more forward peaked. This indicates the growing importance of the $t$-channel 
exchange terms with increasing beam momenta. 
\begin{figure}[t]
\centering
\includegraphics[width=.50\textwidth]{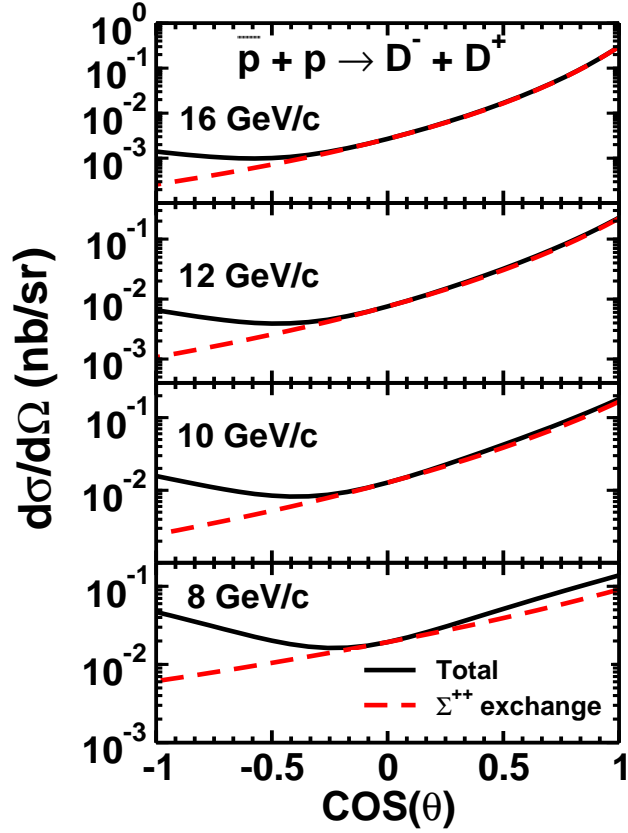}
\caption{(color online)
Differential cross section for the ${\bar p} p \to  D^- D^+$ reaction for
antiproton beam momenta of 8, 10, 12 and 16 GeV/c. The dashed lines 
show the contributions of the $\Sigma_c^{++}$ baryon exchange terms only 
while the solid line represents the cross sections where the amplitudes of 
$\Sigma_c^{++}$ baryon exchange and $\Psi(3770)$ resonance terms are coherently 
added. 
}
\label{fig:Fig10}
\end{figure}

On the other hand, in the differential cross sections of the ${\bar p} p \to  D^- 
D^+$ reaction, the interference effects of the $t$-channel $\Sigma_c^{++}$ baryon 
exchange and the $s$-channel $\Psi(3770)$ resonance terms are visible even at 
higher beam momenta, as can be seen in Fig.~10. In this figure the dashed lines 
represent the cross sections when the contributions of only the $\Sigma_c^{++}$ 
baryon exchange terms are included in the amplitude, while the solid lines
show the results where the amplitudes of $\Sigma_c^{++}$ baryon exchange and 
$\Psi(3770)$ resonance terms are coherently added in the cross sections. One 
notices that the inclusion of the $\Psi(3770)$ resonance terms changes the cross 
sections drastically at the backward angles. This effect is visible even at the  
beam momentum of 16 GeV/c. Thus the measurements of the angular distributions of 
the ${\bar p} p \to  D^- D^+$ reaction even at higher beam momenta can provide 
signals for the $\Psi(3770)$ resonance. Such a study would be  complimentary 
to the methods proposed in Ref.~\cite{xia13}

It should be mentioned here that the physics of the charmed ${\bar D}D$-meson 
production in ${\bar p}p$ annihilation for beam momenta in excess of 3 GeV 
may also involve vector resonances other than $\Psi(3770)$. Some of these 
resonances are $J/\Psi$, $\Psi(2S)$, $\Psi(4040)$, and $\Psi(4160)$. However, 
only $\Psi(3770)$ whose mass is just above the ${\bar D}D$ production threshold, 
has a substantial branching ratio (about 93$\%$) for decay into the ${\bar D}D$ 
channel~\cite{oli14}. Masses of both  $J/\Psi$ and $\Psi(2S)$ are below the ${\bar 
D}D$ production thresholds and their widths are about 2 orders of magnitude smaller 
than that of $\Psi(3770)$. Therefore, their decay to the ${\bar D}D$ channel 
is not possible even from the higher ends of their mass spectrum. Nevertheless, 
because the mass of $\Psi(2S)$ (3.686 GeV) is just below the ${\bar D}D$ 
threshold, it may possibly decay to this channel due to off-shell effects.
However, no branching ratio is known for this decay mode as per the latest PDG 
compilation~\cite{oli14}. Therefore, we have not considered this resonance 
in our work. 

On the other hand, the masses of the resonances $\Psi(4040)$ and $\Psi(4160)$ are
well above the ${\bar D}D$ threshold, and therefore they can decay to the 
${\bar D}D$ channel. However, the branching ratios for these decays are hardly 
quotable  according to the latest PDG compilation. Nevertheless, in Ref.
\cite{hli10} the branching ratios for the decays $\Psi(4040) \to {\bar D}D$ and 
$\Psi(4160) \to {\bar D}D$ have been estimated by fitting to the $e^+e^- \to 
{\bar D}D$ data of the Belle Collaboration~\cite{pak08} in the invariant mass 
region of $3.8-4.3$ GeV. In this procedure these resonances are parametrized in 
terms of the Breit-Wigner form. The estimated branching ratios for the two decay 
channels are found to be (25.3 $\pm$ 4.5)$\%$ and (2.8 $\pm$ 1.8)$\%$, respectively. 
These can be used to obtain the coupling constants $g_{\Psi(4040){\bar D}D}$ and 
$g_{\Psi(4160){\bar D}D}$. However, to calculate the cross sections for ${\bar p} 
+ p \to \Psi(4040) \to {\bar D} D$ and ${\bar p} + p \to \Psi(4160) \to {\bar D} D$
processes, we also require the coupling constants  $g_{\Psi(4040) {\bar p} p}$
and  $g_{\Psi(4160) {\bar p} p}$, about which no information is available.
In any case, we performed calculations for the cross sections of these reactions
by taking the values of the  $g_{\Psi(4040) {\bar p} p}$ and $g_{\Psi(4160) {\bar 
p} p}$ to be the same as that of $g_{\Psi(3770) {\bar p} p}$. We find that the
resulting cross sections in the relevant region are much smaller in comparison
to the total cross sections shown in Figs.~7 and 8. Therefore, the inclusion 
of resonances $\Psi(4040)$ and $\Psi(4160)$ hardly lead to any noticeable change
in the overall conclusions of this paper.

Finally, we discuss the uncertainties and the range of the validity of our results.
The theoretical approach (the effective Lagrangian model or ELM) considered in this
work has mesons and baryons as effective degrees of freedom. This model will 
be valid in the energy range where consideration of explicit quark degrees of 
freedom is not required. Therefore, our model is certainly applicable in the range of 
beam momenta (from threshold to 20 GeV/c) considered in this work. However, the 
calculations performed within this model are sensitively dependent on the values of 
the coupling constants at various vertices involved in the $t$-channel and $s$-channel 
diagrams, on the shape of the form factor and the value of the cutoff parameter 
involved therein, and on the parameters involved in the initial- and final-state 
interaction scattering matrices. The extents of uncertainty in our results due to all 
these issues are discussed in the following. 

We have taken the coupling constants (CCs) at the vertices involved in the $t$-channel 
diagrams from Refs.~\cite{hai11a,hai07,hai08} where they have been fixed by using the 
SU(4) symmetry arguments in the description of the exclusive charmed hadron 
production in the ${\bar D}N$ and $DN$ scattering within a one-boson-exchange picture. 
The same coupling constants were used in the description of the charmed hadron 
production within the J\"ulich meson-exchange model in Refs.~\cite{hai10,hai15}. 
Furthermore, these coupling constants were also used in Ref.~\cite{hob14} to 
investigate the role of intrinsic charm in the nucleon using a phenomenological model 
formulated in terms of effective meson-baryon degrees of freedom. Thus, the coupling 
constants used in the calculations of the $t$-channel diagrams of our model are 
quite standard. The CCs at the vertices involved in the $s$-channel diagrams are 
determined from the experimentally determined branching ratios of the decay of the 
$\Psi(3770)$ resonance into the relevant channels. Therefore, uncertainties in our 
cross sections due to the coupling constants are minimal.

There may indeed be some uncertainty in our cross sections coming from the shape
of the form factor [and the value of the cutoff parameter ($\lambda_i$) involved 
therein] that are used to regulate the off-shell behavior of various vertices. As 
stated above, we have employed a monopole form factor as given by Eq.~3, with a 
$\lambda_i$ of 3.0 GeV. A form factor of a different shape and/or a different value of 
the cutoff parameter would lead to a different cross section.  For example, using a 
quadrupole form factor [of the type given by Eq.~ 4] with the same cutoff parameter 
leads to enhancement in the cross sections by factors of $3-4$. Changing $\lambda_i$ 
from 3 to 3.5 increases the cross sections by a factor of up to 2. We have tried to 
minimize these uncertainties in the cross sections by using the same shape (monopole) 
of the form factor and the same value of $\lambda_i$ that were used in our previous 
study of the charmed baryon production~\cite{shy14}. As in other cases,  issues 
related to form factor will be finally settled once the data become available 
on the charmed meson and baryon production in ${\bar p}p$ annihilation from the 
${\bar P}ANDA$ experiment.

The initial- and final-state interactions, which are the important ingredients of our 
model, provide another source of uncertainty in our results. We treat these effects
within an eikonal-approximation-based phenomenological method. Generally, the 
parameters of this model are constrained by fitting to the experimental data. 
Because of the lack of any experimental information, it is not yet possible to test
our model thoroughly. The absolute magnitude of our cross sections may have some
uncertainties due to this. Nevertheless, in our study we have used the same set of 
distortion parameters that were used in our previous calculations of the charmed 
baryon production in the same reaction. These parameters reproduce the data for the 
${\bar \Lambda} \Lambda$ channel and the cross sections for the ${\bar \Lambda}_c 
\Lambda_c$ channel calculated within the J\"ulich meson-exchange model where distortion 
effects are treated more rigorously within a coupled-channels approach. Therefore, 
the initial and final channel distortion effects included in our model are checked 
against the other independent sources. 
  
\section{Summary and conclusions}

In summary, we studied the ${\bar p} + p \to {\bar D}^0 D^0$ and ${\bar p} + p \to 
 D^- D^+$ reactions by using a single-channel effective Lagrangian model that
involves the meson-baryon degrees of freedom. The dynamics of the production 
process has been described by the $t$-channel $\Lambda_c^+$, $\Sigma_c^+$ and 
$\Sigma_c^{++}$ baryon exchange diagrams and also the $s$-channel excitation, 
propagation and decay of the $\Psi(3770)$ resonance. The initial- and final-state 
interactions have been accounted for by an eikonal type of phenomenological model. 
The coupling constants at the baryon exchange vertices were taken from Refs.
~\cite{hai11a,hai07,hai08}, which were the same as those used in the study of the 
${\bar p} + p \to {\bar \Lambda}_c^ - \Lambda_c^+$ reaction at the similar vertices 
in Ref.~\cite{shy14}.  The CCs at $\Psi {\bar p} p$, $\Psi {\bar D}^0 D^0$ and 
$\Psi D^- D^+$ vertices have been determined from the branching ratios for the decay 
of $\Psi(3770)$ resonance into the relevant channels as given in Refs.~\cite{abl06} 
and~\cite{abl14}. The off-shell corrections at various vertices have been accounted 
for by introducing monopole form factors with a cutoff parameter of 3.0 GeV. The 
same form factor with the same value of the cutoff parameter was also used in our 
study of the ${\bar p} + p \to {\bar \Lambda}_c^- \Lambda_c^+$ reaction~\cite{shy14}. 
The parameters involved in the initial-state interaction scattering matrices were 
also taken to be the same as those used in Ref.~\cite{shy14}.

Since the cross sections of the ${\bar p} + p \to D^- D^+$ reaction are strongly 
suppressed due the smaller coupling constants of the vertices involving 
$\Sigma_c^{++}$ baryon exchange, the inclusion of the $\Psi(3770)$ resonance 
produces a sizable enhancement in the ${\bar p} + p \to D^- D^+$ cross sections 
around the resonance energy. However, their effect is not so strong in case of the 
${\bar p} + p \to {\bar D}^0 D^0$ reaction where the baryon exchange cross sections 
are quite large $-$ they vary between $100-400$ nb for antiproton beam momenta 
between $6.4-6.8$ GeV/c. Therefore, the inclusion of the $\Psi(3770)$ resonance in this 
case produces only a small kink in the total cross section near the resonance 
energy.

On the other hand, the differential cross sections for both the reactions are 
affected in a major way by the $\Psi(3770)$ resonance contributions for antiproton 
beam momentum near the resonance peak. In case of the ${\bar p} + p \to {\bar D}^0 
D^0$ reaction, the $\Psi$ resonance contributions introduce sizable reduction 
(enhancement) in the DCS at the backward (forward) angles. For the ${\bar p} + p \to 
D^- D^+$ reaction, the shape of the DCS changes drastically by the inclusion of 
the $\Psi$ resonance term $-$ it changes to a $p$-wave type of distribution from a 
$s$-wave shape.  This drastic shape change of the DCS can perhaps be exploited in 
a dedicated experiment at the ${\bar P}ANDA$ facility to pin down the 
$\Psi(3770)$ resonance.

At higher antiproton momenta, the total cross section of the ${\bar p} + p \to 
{\bar D}^0 D^0$ reaction is dominated by the contributions of the $\Lambda_c^+$ 
baryon exchange. The cross section peaks around $p_{{\bar p}}^{lab}$ of 9 GeV/c. At 
a $p_{{\bar p}}^{lab}$ of 15 GeV/c, which is of interest to the ${\bar P}ANDA$ 
experiment, the total cross section of this reaction is about 550 nb which is at
least 5 times larger than the largest value of this cross section reported 
previously. Of course, previous calculations have used different types of models 
that invoke explicitly the quark degrees of freedom in their calculations, which 
may make them more adequate for energies higher than those of the ${\bar P}ANDA$ 
experiment. Therefore, it is not trivial to understand the reasons for the large 
difference seen between their cross sections and ours. The future ${\bar P}ANDA$ 
experiments at FAIR are expected to clarify the situation.

Within our model the cross sections of the ${\bar p} + p \to D^- D^+$ reaction are
strongly suppressed as compared to those of the ${\bar p} + p \to {\bar D}^0 D^0$ 
reaction. This is due to the fact that the latter is dominated by the $
\Lambda_c^{+}$ baryon exchange mechanism, while the former gets a contribution only 
from the $\Sigma_c^{++}$ exchange whose couplings are much lower than those of the 
$\Lambda_c^{+}$ exchange vertices. However, in the coupled-channels meson-exchange
model of Ref.~\cite{hai14}, the ${\bar p} + p \to D^- D^+$ cross sections are 
even larger than the ${\bar p} + p \to {\bar D}^0 D^0$ ones. This is a result of the 
coupled-channels treatment in the incident channel which accounts effectively for 
two-step inelastic processes involving $\Lambda^+_c$ "baryon exchange." 

The differential cross sections of the ${\bar p} + p \to {\bar D}^0 D^0$ reaction 
at higher values of $p_{{\bar p}}^{lab}$ are strongly forward peaked and are 
so strongly dominated by the contributions of the $\Lambda_c^+$ baryon exchange
terms that the interference terms of various other contributions ($\Sigma_c^+$ 
baryon exchange and $\Psi$ resonance) become insignificant. However, in the case
of the ${\bar p} + p \to D^- D^+$ reaction, differential cross sections have 
significant contributions from the interference terms of $\Sigma_c^{++}$ baryon 
exchange and the $\Psi$ resonance process even at higher antiproton beam momenta. In 
view of these results, it should be possible to pin down the $\Psi(3770)$ resonance 
contributions in these reactions in dedicated experiments in the relevant antiproton 
energy regions.  
   
\section{acknowledgments}
 
This work has been supported by the Deutsche Forschungsgemeinshaft (DFG) under Grant
No. Le439/8-2 and Helmholtz International Center (HIC) for FAIR and the Council of 
Scientific and Industrial Research (CSIR), India.

\end{document}